\documentclass[journal]{journal}
\usepackage{cite}
\usepackage[pdftex]{graphicx}
\usepackage{xcolor}
\usepackage[cmex10]{amsmath}
\usepackage{array}
\usepackage{mdwmath}
\usepackage{mdwtab}
\usepackage{multirow}
\usepackage[font=footnotesize]{subfig}
\usepackage{fixltx2e}
\usepackage{subcaption}
\usepackage{float}

\begin{document}
\title{It’s My Turn! Interpretable Methods Reveal \\Developmental Differences from \\Mother-Child Conversational Dynamics
}
\author{A'di Dust, Anura Deshpande, Pat Levitt, Maja Matari\'c}

\maketitle
\thispagestyle{empty}

\begin{abstract}
Understanding conversational interactions between mothers and children is an important challenge for computational behavioral modeling. Existing methods use traditional statistical approaches and subjective measures of conversational dynamics, without incorporating temporal or probabilistic modeling. We developed an interpretable computational framework that combines sequence analysis, information-theoretic measures, and Bayesian hierarchical modeling to characterize temporal conversational dynamics in parent-child play. We applied this framework to interactions from 94 mother-child dyads with two-year-old children. Sequence analysis revealed that dyads with lower language development (as defined by scores on the Bayley developmental test) exhibited more heterogeneous conversational organization and fewer serve-and-return exchanges than higher-development dyads, a reciprocal \textit{turn-taking} pattern in which one partner initiates communication and the other responds. Mutual information analyses suggested differences in conversational predictability between lower- and higher-development dyads, although the bootstrap confidence intervals overlapped. Bayesian hierarchical models estimated temporal trajectories for maternal consecutive turns, child responses following maternal speech, and overlapping responses; HDIs showed that trajectories differed in whether they excluded 0, but there were no statistically credible differences between the Bayley groups. Overall, these results indicate that children with higher developmental scores have more organized and predictable communication styles, with more ``serve-and-return" patterns with their mothers. These results demonstrate that interpretable computational representations of conversational dynamics capture developmental differences beyond simple measures of speech quantity.
\end{abstract}


\IEEEpeerreviewmaketitle

\section{Introduction}
The quality of communication between mothers and their children is vital for child development \cite{abah2025role, chen2023mother, baumwell1997maternal, mcgillion2013supporting, solomon1996defining}. Interactions during play are particularly informative, as they have been shown to predict child temperament and other developmental outcomes \cite{runcan2012importance}. Not only is play important, but its positive qualities can also be coached. In interventions such as the Filming Interactions to Nurture Development (FIND), insights into the most important aspects of play are taught through video-based coaching. This has been shown to improve caregiver-child interactions, with measurable benefits demonstrated in randomized controlled trials \cite{liu2025improving, fisher2006effects, fisher2016promoting}. With these new methods, it is critical to fully understand the qualities of play that affect development to inform future coaching endeavors.

To investigate whether developmental differences are associated with changes in \textit{conversational structure}, we modeled mother-child conversations as sequences of speaker transitions and analyzed turn-taking. We use three complementary methods: sequence analysis to characterize transition patterns, mutual information to quantify the predictability of speaker transitions, and hierarchical Bayesian modeling to examine how \textit{turn-taking} evolves over the course of an interaction.

Our findings indicate that children with higher language and cognitive proficiency have more organized communication styles than those with lower proficiency. Specifically, higher-proficiency children tend to engage in more serve-and-return-style communication, a concept in developmental science that refers to the back-and-forth nature of mother-child communication \cite{shonkoff2008national}. Children with high levels of cognitive and language proficiency also exhibit more predictable, structured patterns of turn-taking. Finally, although we found no statistically credible differences between children with higher and lower cognitive or language scores in our Bayesian analyses, conversational dynamics changed over time in ways that varied by developmental level, as defined by Bayley developmental test scores. Children with lower developmental scores exhibited increasingly child-led participation, whereas children with higher language scores increasingly produced overlapping responses following maternal turns, with mothers becoming less likely to produce consecutive turns. These findings suggest that in play-based interactions, mothers may benefit from focusing on serve-and-return-style communication and organized, structured interactions, while also limiting interruptions to their 2-year-old children in play, especially as a conversation continues.

\section{Background and Related Work}
Extensive literature has shown remarkable relationships between the type and quality of mother-child interaction and child outcomes. This work focuses on turn-taking during interactions because turn-taking is a fundamental component of human communication that enables coordinated social interaction and supports later language acquisition \cite{stevanovicExperienceSharingEmotional2015, kellyItsYourTurn2022, gilkersonLanguageExperienceSecond2018, endevelt-shapiraMotherInfantSocial2024, nguyenYourTurnMy2023}. Prior work has operationalized turn-taking using measures such as response latency, overlap duration, interruption frequency, contingency rates, and conversational balance \cite{cosperHumanTurntakingDevelopment2025}. Researchers have argued that the ability to participate in structured exchanges emerges before children acquire substantial lexical or grammatical knowledge, suggesting that conversational organization develops independently of language content \cite{gratierEarlyDevelopmentTurntaking2015, cosperHumanTurntakingDevelopment2025}. Consistent with this view, turn-taking behaviors are observed early in infancy and exhibit similarities across languages and cultures \cite{stiversUniversalsCulturalVariation2009}. 

Human conversation and protoconversation---the prelinguistic exchange of communicative signals before the emergence of lexical and grammatical structure---are both organized around the sequential exchange of turns between interaction partners \cite{yooOriginProtoconversationExamination2018}. Between ages 0 and 3, children pause frequently between vocalizations and maintain a serve-and-return pattern of vocal interaction \cite{bloomTurnTakingAffects1987}. These findings suggest that early conversational coordination may contribute to later language development rather than simply reflecting emerging linguistic ability. 

In early conversation, when children are younger than 24 months, children more frequently interrupt their mothers and show less coordination between their vocalizations and their mothers' \cite{gratierEarlyDevelopmentTurntaking2015}. However, by 24 months, turn-taking becomes more fluent and adult-like. At that age, turn-taking is fluid even with delays, irrelevant responses, and nonresponse \cite{kellyItsYourTurn2022}. In addition, children also end overlapping turns at a higher rate than their mothers \cite{cosperHumanTurntakingDevelopment2025}. The vocalizations themselves also become more similar by the end of a conversation than at the beginning, showing an increase in fluency on a conversation-wide turn scale \cite{li2022multi}.

Together, these findings suggest that turn-taking develops from early caregiver-supported protoconversation to increasingly coordinated, contingent turn-taking in the first two years of life. However, despite extensive descriptive work documenting developmental changes in turn-taking, relatively little research has developed interpretable modeling that captures multiple dimensions of turn-taking behavior simultaneously. Furthermore, less is known about how turn-taking patterns vary among children with different language and cognitive abilities at approximately 24 months of age, where turn-taking becomes more contingent and fluent. Zhang et al. \cite{zhang2025turn} found that there are some differences in which typically developing dyads are more entrained in a way that was sequentially dependent compared to atypically developing children. However, little work has been done comparing children at different ends of the typically developing spectrum, particularly regarding turn-taking sequences. Our methods provide insight into this understudied area and demonstrate that machine learning and Bayesian approaches are important for studying highly heterogeneous populations.

\section{Dataset}
This work included 94 mother-child dyads from a larger, ongoing, longitudinal study of 340 dyads \cite{gateau2026evaluating}. The children took the Bayley-3 developmental assessment with a trained research assistant in either English or Spanish and a free play session with their mother as part of a larger study. Children were considered "Low" if their Bayley score was below 90, "High" if their score was greater than 110, and "Average" otherwise. On the language scale, 39 children were Low, 28 were Average, and 27 were High. On the cognitive scale, 10 children were Low, 58 were Average, and 26 were High. 

Inclusion criteria were completion of all Bayley categories at age 2 and a 3-minute ± 15-second free-play interaction video, during which the mother and child spoke only English and/or Spanish. The selected videos totaled 4 hours and 52 minutes of recorded interactions over 94 dyads. 

\begin{figure}
    \centering
    \includegraphics[width=0.85\linewidth]{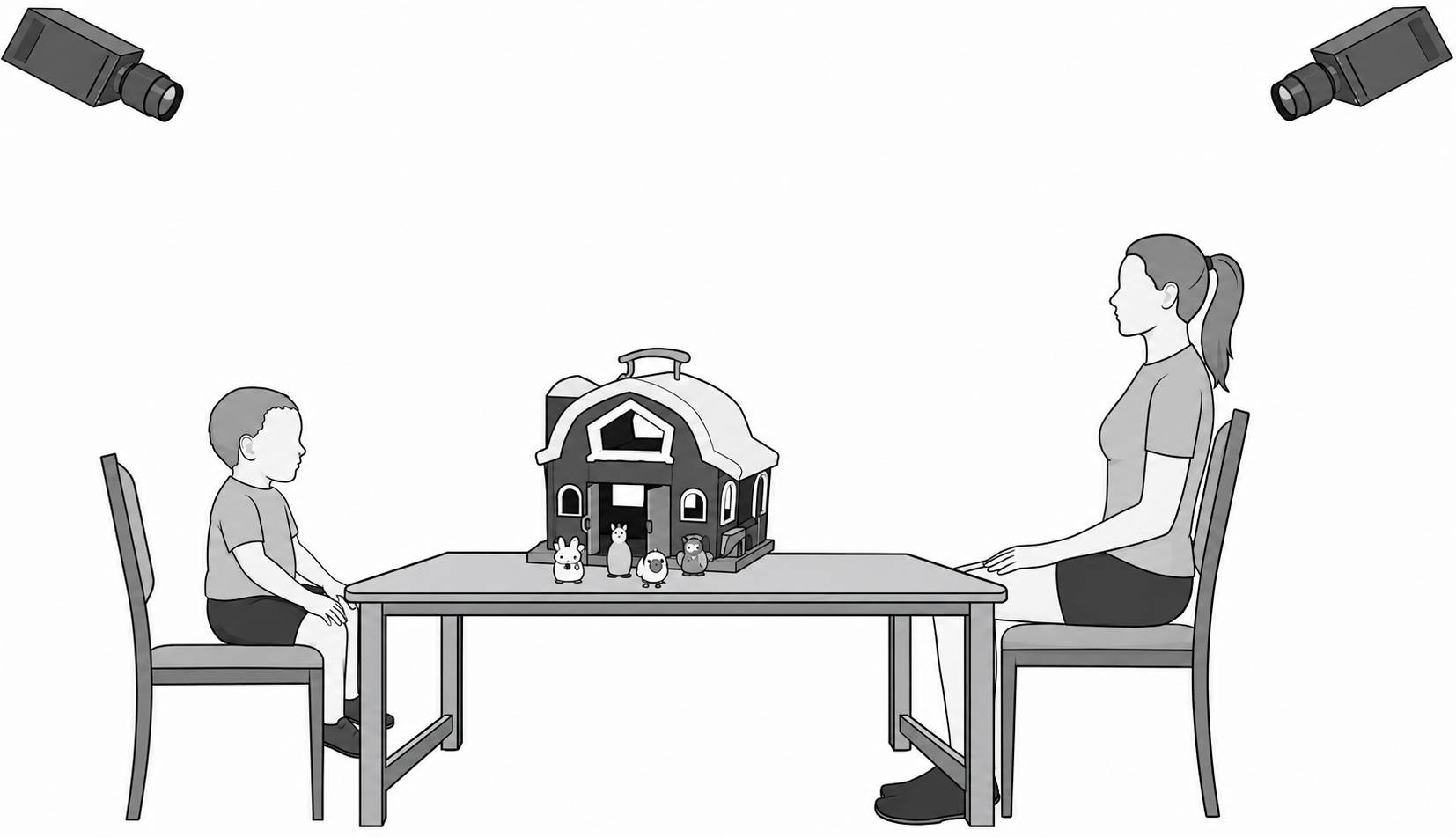}
    \caption{The setup of mother-child free play interaction.}
    \label{fig:setup}
\end{figure}

The mother-child free play interaction involved a short play session in which the mother was instructed to play with her child as she would at home. The dyad was seated at a small table and given a toy farmhouse with toy animals and a shape-block puzzle. Two Noldus PTZ corner-mounted cameras with Viso Software for high-quality audio-video recording were positioned in opposite corners of the room: one directed toward the mother’s face and body, and the other toward the child's. This setup is shown in Figure \ref{fig:setup}. 

To label turn-taking behaviors, each video was manually diarized, as automatic methods using open-source models were unable to accurately identify speakers, likely due to large, non-stationary noise artifacts from children dropping toys, and to significant age differences between the training sets and the young children in our study. One researcher annotated 94 videos. The annotator used Label Studio \cite{LabelStudio} to segment and label utterances from each video. Each utterance, whether verbal or nonverbal/pre-verbal, was annotated as adult, child, or overlapping speakers. Utterances by the same speaker with more than 500 ms of silence between them were considered separate utterances. To assess annotation reliability, another researcher independently annotated a random subset of 10 videos. Inter-rater agreement, measured by Needleman-Wunsch identity, averaged 82.1\%. Disagreements primarily concerned whether adjacent vocalizations were marked as a single turn or multiple turns.

The dyads included in this analysis were from a large city in the United States and had a diverse demographic makeup. 52\% of the children were female. A plurality of children were White (43\%), followed by Black (15\%), Asian (9\%), Native American (7\%), Native Hawaiian or other Pacific Islander ($<$1\%), with the remaining 25\% of mothers not wishing to disclose their child's race. The cohort was mostly Hispanic or Latino, with 71\% identifying as Hispanic or Latino, 26\% not identifying as Hispanic or Latino, and 3\% not wishing to disclose their ethnicity. The main language spoken in the participants' households was English only (57\%), followed by both English and Spanish (22\%) and Spanish only (13\%), with the remaining households speaking another language in addition to English. 

\section{Approach}
To investigate whether developmental differences are reflected in mother-child conversational dynamics, we analyzed turn-taking using three complementary, interpretable methods. We first used sequence analysis to characterize the structure of conversational sequences. Then, we quantified the predictability of speaker transitions using mutual information. Finally, we modeled how transition probabilities changed over time using a hierarchical Bayesian framework. Together, these methods capture differences in interaction structure, predictability, and temporal dynamics across developmental groups.

\subsection{Sequence Analysis}

The first interpretable model of turn-taking we implemented analyzed turn-level sequences. We used simple statistical analysis and the Needleman-Wunsch (NW) sequence similarity algorithm, a method developed for comparing DNA sequences \cite{needlemanGeneralMethodApplicable1970}. While our NW model used full turn-taking sequences, composed of the speaker label (mother, child, or an overlap), our statistical analyses relied on 3-turn sequences, since most turn-taking formats in 24-month-old children involve 3 turns \cite{gratierEarlyDevelopmentTurntaking2015}.

Our NW implementation took speaker turns and mapped them into a 2-dimensional grid space, with the current speaker on one axis (x-axis) and the next speaker on the other (y-axis). This approach has been shown to promote understanding of longitudinal inter- and intra-dyad behavioral transitions as in \cite{brinbergAnalyzingDyadicData2018}. Our grid-based approach begins by turning the 2D transitions into “words”, or sequences. For example, the transition (mother → child → child) would be transcribed as “ACC”. The entire sequence for each dyad is transcribed this way. 

To measure the similarity between different dyads, we calculated the Needleman-Wunsch Similarity. Let the first sequence be $x_1, x_2, …, x_m$ where each $x_i$ is the speaker at turn $i$, and similarly let the second sequence be $y_1, y_2, …, y_n$. 

We define a scoring function $S$ that returns scores for matches, mismatches, and gaps. A match occurs when the speakers at aligned positions are the same, giving a score of +1. For example, $S(C,C) = +1$ if the aligned turns are both from child speakers. Mismatches occur when the aligned speakers differ and incur a penalty of -1. Finally, gaps correspond to insertions or deletions introduced during the alignment and incur a penalty of -0.5.

Using this scoring scheme, we defined a matching matrix $M$ of size
$(m+1)\times(n+1)$. Each entry $M(i,j)$ represents the score of the
optimal alignment of the prefixes $x_1,\ldots,x_i$ and
$y_1,\ldots,y_j$. The matrix is initialized with
$M(0,0)=0$, $M(i,0)=-0.5i$, and $M(0,j)=-0.5j$. The remaining entries are computed
recursively as 
\begin{equation}
    M(i,j) =
    \max \Bigg\{
    \begin{aligned}
    &M(i-1,j-1) + S(x_i,y_j),\\
    &M(i-1,j) + S(x_i,\_),\\
    &M(i,j-1) + S(\_,y_j)
    \end{aligned}
    \Bigg\}
\end{equation}

The final similarity score is given by $M(m,n)$, which represents the optimal alignment between the two sequences.

\subsection{Mutual Information}

Mutual information quantifies the dependence of one variable on another. We used mutual information to measure how much information the current speaker's identity provides about the next speaker's identity. If speaker identities are independent across turns, the mutual information is zero; higher values indicate that knowledge of the current speaker increases the predictability of the next speaker.

For each interaction, conversational turns were coded as adult (a), child (c), or overlap (o). Consecutive turns were sorted chronologically within each dyad. Each transition was represented as a pair $(s_t, s_{t+1})$, where $s_t$ denotes the variable current speaker identity and $s_{t+1}$ denotes the next speaker identity. Their possible values, denoted by $s_t$ and $s_{t+1}$, correspond to the three speaker identities $\{\text{adult}, \text{child}, \text{overlap}\}$. Mutual information between adjacent speaker identities was computed as:

\begin{equation}
    I(s_t;s_{t+1}) = \sum_{s_t}\sum_{s_{t+1}}p(s_t, s_{t+1})log_2\frac{p(s_t, s_{t+1})}{p(s_t)p(s_{t+1})}
\end{equation}

Here, $p(s_t, s_{t+1})$ denotes the joint probability of observing a particular speaker transition, while $p(s_t)$ and $p(s_{t+1})$ denote the marginal probabilities of the current and next speaker identities, respectively.
To facilitate interpretation, we additionally computed the entropy of the next speaker

\begin{equation}
    H(S_{t+1}) = -\sum_{s_{t+1}}p(s_{t+1})log_2p(s_{t+1})
\end{equation}

and the conditional entropy of the next speaker given the current speaker:

\begin{equation}
    H(s_{t+1}|S_t) = H(s_{t+1})-I(s_t;s_{t+1})
\end{equation}

The proportion of uncertainty reduced by observing the current speaker was calculated as:

\begin{equation}
    Uncertainty Reduction = \frac{I(s_t;s_{t+1})}{H(s_{t+1})}\times 100
\end{equation}

This metric allows mutual information to be interpreted as the percentage reduction in uncertainty about the next speaker after observing the current speaker.

To determine whether developmental differences affect turn-taking, mutual information was computed separately for children classified as High, Average, and Low according to Bayley Language and Cognitive composite categories. All transitions from dyads belonging to a given category were pooled to estimate the category-level transition mutual information.

\subsection{Hierarchical Bayesian Modeling}

To model possible time-variant aspects of turn-taking, we used a Bayesian hierarchical model. For each dyad $d$, let $prevobs_{d,t}=obs_{d, t-1}$ and scale the turn number to $\tau_{d,t}\in [-1,1]$. For each transition type, $k \rightarrow j$, we model a dyad-specific intercept and slope. At the population level, the intercept and slopes have hierarchical priors:

\begin{align}
\alpha_{\mu,k,j} &\sim \mathcal{N}(0,1), &
\sigma_{\alpha,k,j} &\sim \operatorname{HalfNormal}(0.5)
\end{align}

\begin{align}
\beta_{\mu,k,j} &\sim \mathcal{N}(0,0.5), &
\sigma_{\beta,k,j} &\sim \operatorname{HalfNormal}(0.5).
\end{align}

Dyad-level deviations are 

\begin{equation}
    \alpha_{d,k,j}
    =
    \alpha_{\mu,k,j}
    +
    \alpha^{\mathrm{raw}}_{d,k,j}\sigma_{\alpha,k,j}
\end{equation}

where $\alpha^{\mathrm{raw}}_{d,k,j}\sim\mathcal N(0,1).$

To analyze time effects across the different Bayley groups, we include both dyad-level variation and group effects in the time slope, with indicators for the high and low Bayley groups.

\begin{align}
    \beta_{d,k,j}
    &= \beta_{\mu,k,j}
    + \beta_{d,k,j}^{\mathrm{raw}}\,\sigma_{\beta,k,j} \\
    &\quad + \mathbf{1}(high_d)\,\beta_{\mathrm{high},k,j}
    + \mathbf{1}(low_d)\,\beta_{\mathrm{low},k,j} \nonumber
\end{align}

 where $\beta^{\mathrm{raw}}_{d,k,j}\sim\mathcal N(0,1)$, with

\begin{align}
    \beta_{high,k,j} &\sim \mathcal{N}(0, 0.5), &
    \beta_{low,k,j} &\sim \mathcal{N}(0,0.5)
\end{align}

The log-odds associated with the transition $k \rightarrow j$ at turn $t$ for dyad $d$ are

\begin{equation}
\eta_{d,t,k,j}=\alpha_{d,k,j}+\beta_{d,k,j}\tau_{d,t}
\end{equation}

To formulate the multinomial model, we use the “overlap” transition category as the reference level and set its logit to 0, such that:

\begin{equation}
    logits_{d,t,k}=(\eta_{d,t,k,1}, …, \eta_{d,t,k,J-1},0)
\end{equation}

where $J$ denotes the number of transition categories. The next transition is predicted from the previous observation and the time-varying transition probabilities:

\begin{equation}
    y_{d,t} \sim Categorical(p = \mathrm{softmax}({d,t,prevobs_{d,t}}))
\end{equation}

where the logits indexed by $(d,t,\mathrm{prevobs}_{d,t})$ are transformed by the softmax to obtain $p$. We fit the model using 4 chains, 2000 tuning samples, 2000 draws, and a target acceptance of 0.95, because these parameters were sufficiently stable with no divergences.

The resulting fit model was well-sampled. For each chain, the Bayesian Fraction of Missing Information (BFMI) is near 1, much higher than the minimum 0.3, indicating sufficient exploration of the posterior distribution. For all variables, there were no divergent transitions and $\hat{R}=1$ for all variables. This model was created using the \textit{PyMC 6.3} package in Python, using NUTS for the free parameters \cite{pymc2023}. A visualization of the hierarchical model is provided in Appendix \ref{app:bayesian}.

\section{Results}
Sequence analysis characterized differences in interaction structure across developmental groups, mutual information quantified the predictability of speaker transitions, and hierarchical Bayesian modeling allowed us to examine changes in conversational dynamics over time. We present our results of the three analyses below. 

\subsection{Sequence Analysis}

To analyze the order of turn-taking sequences, we created linear mixed-effects models to predict similarity between dyadic sequences, with fixed effects for group comparisons (e.g., low Bayley vs. high Bayley). We also included random effects for each dyad in the score-similarity comparison. The linear mixed-effects models both had 4371 observations and a mean group size of 47 because we do not duplicate comparisons by order. We used REML for fitting the model. Both the language and models converged successfully.

The model included random intercepts for the two dyads in each pairwise comparison (dyad 1 and dyad 2), with estimated variance components of 43.85 and 61.20, respectively. Fixed-effect contrasts showed significant differences only for comparisons involving the low Bayley group. Those estimates were below the Average-Average reference, indicating lower predicted sequence similarity.

\begin{figure}[t]
    \centering

    \subfloat[Language]{%
        \includegraphics[width=0.95\columnwidth]{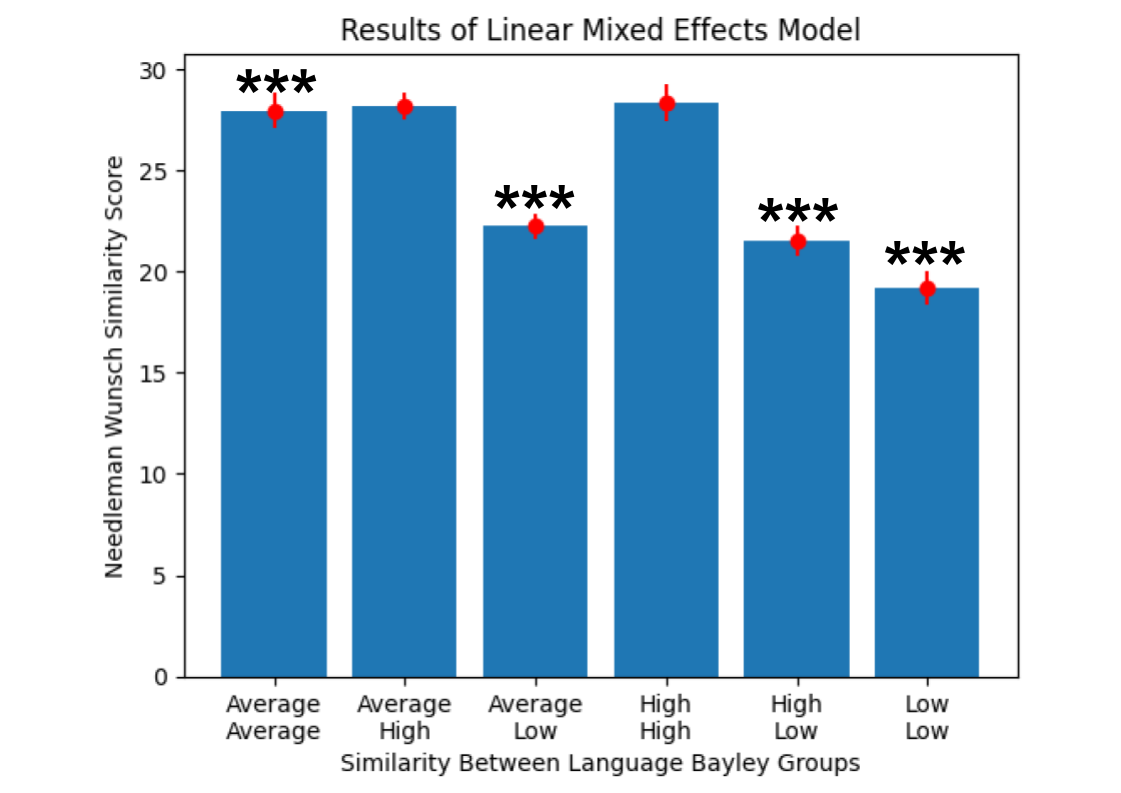}
    }

    \par\vspace{0.3cm}

    \subfloat[Cognitive]{%
        \includegraphics[width=0.95\columnwidth]{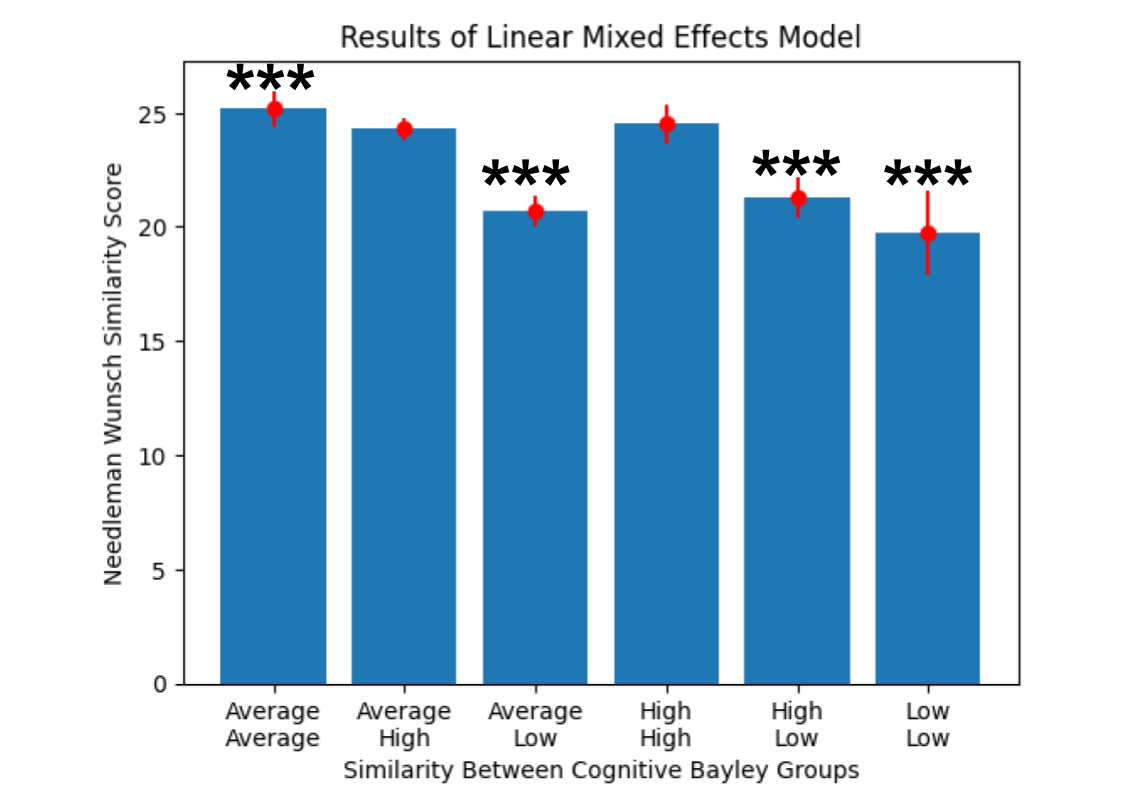}
    }

    \caption{Linear mixed-effects estimates of NW similarity by Bayley group. Plotted values equal the Average-Average intercept plus the corresponding coefficient; *** ($p < 0.001$).}
    \label{fig:mixed_effects}
\end{figure}

The cognitive Bayley model showed a similar pattern. Estimated random-intercept variance components for the two dyad roles (dyad 1 and dyad 2) were 46.17 and 63.92, respectively. Fixed-effect contrasts showed significantly lower sequence similarity for comparisons involving the low Bayley group relative to the Average-Average reference; no other group contrasts were significant. Figure \ref{fig:mixed_effects} shows the fixed-effect estimates relative to this reference.

In addition to performing NW linear mixed-effects analysis, we analyzed 3-turn sequences from the Bayley group to identify significant differences in common turn-taking patterns. Although adult conversations often revolve around 5-turn sequences \cite{solomonUsingSequenceAnalysis2022}, early and protoconversations do not exhibit this pattern, with low frequencies of specific 5-turn sequences, motivating our use of 3 turns. Significant differences for language Bayley are in Table \ref{tab:language_anova}. All sequence ANOVAs, including non-significant comparisons, are available in the appendix.

Of the 27 possible three-turn sequences, three differed significantly across language Bayley groups after Holm adjustment; none differed significantly across cognitive Bayley groups after adjustment. The tables report Holm-adjusted values as \(p\). All three significant language results had large effect sizes. Descriptively, lower language Bayley scores were associated with more adult-dominated AAA turn-taking, while ACA and CAC turn-taking sequences increased in frequency across the language Bayley groups.

\begin{table*}[t]

\centering
\small
\begin{tabular}{|l|l|l|ccc|}
\hline
\multirow{2}{*}{\textbf{Sequence}} &
\multirow{2}{*}{\textbf{ANOVA}} &
\multirow{2}{*}{\textbf{Effect Size}} &
\multicolumn{3}{c|}{\textbf{Average Frequency by Bayley}} \\ \cline{4-6}
& & &
\textbf{Low} &
\textbf{Average} &
\textbf{High} \\ \hline

AAA & F(2, 91) = 7.47, $p=0.025$ & $\eta_p^2=0.14$ (large) & 15.44\% & 9.91\% & 9.76\% \\ \hline
ACA & F(2, 91) = 13.06, $p<0.001$ & $\eta_p^2=0.22$ (large) & 9.01\% & 12.51\% & 13.61\% \\ \hline
CAC & F(2, 91) = 8.41, $p=0.016$ & $\eta_p^2=0.16$ (large) & 4.21\% & 6.24\% & 8.03\% \\ \hline

\end{tabular}
\caption{ANOVA differences in frequency of 3-turn sequences by language Bayley group.}
\label{tab:language_anova}
\end{table*}

\subsection{Mutual Information}

Mutual information was calculated for mother, child, or overlap, based on the described speaker diarization process, across all conversational transitions (n=7510). The mutual information between the current and next speakers was 0.0469 bits. The entropy of the next speaker was 1.2525 bits, while the conditional entropy given the current speaker was 1.2056 bits, corresponding to an uncertainty reduction of 3.74\%.
The transition probability matrix revealed several consistent patterns. All turn types were most likely followed by an adult turn. Child turns were most frequently followed by adult turns 74.3\% of the time, adult turns were followed by another adult turn 58.1\% of the time, and overlap segments were followed by adults 71.8\% of the time. Overall, knowing the current speaker's identity reduced uncertainty about the next speaker's identity by 3.7\%, indicating a relatively weak dependence between adjacent speaker identities across the full dataset.

Always classifying the next speaker as an adult yields a baseline accuracy of 63.7\%, the overall proportion of adult next turns. Because an adult was the most likely next speaker after each current-speaker category, using the current speaker does not change the most likely classification; however, it does shift the estimated probabilities, corresponding to a 3.7\% reduction in uncertainty.

To examine whether conversational structure differed across developmental groups, mutual information was calculated separately for Bayley Language and Cognitive development. For language development, uncertainty reduction increased monotonically from the low (2.79\%) to the average (4.86\%) to the high (5.70\%) groups. A similar pattern was observed for cognitive development, with uncertainty reduction increasing from 2.07\% in low dyads to 3.63\% in average dyads and 5.01\% in high dyads.

This reveals that knowledge of the current speaker reduced uncertainty about the next speaker by approximately twice as much in high-language dyads (5.70\%) as in low-language dyads (2.79\%). Likewise, high-cognitive dyads exhibited more than double the uncertainty reduction observed in low-cognitive dyads.

Across language and cognitive domains, higher Bayley performance was consistently associated with greater mutual information between adjacent speaker identities. These trends suggest that children with high development scores participate in conversations with more structured patterns of speaker exchange. However, the bootstrap 95\% confidence intervals, resampled by dyad 1000 times, shown as error bars in Figure \ref{fig:mi}, showed substantial overlap across developmental groups, indicating that these differences should not be interpreted as statistically significant effects.

\begin{figure*}[t]
    \centering
    \subfloat[Language]{
        \includegraphics[width=0.48\textwidth]{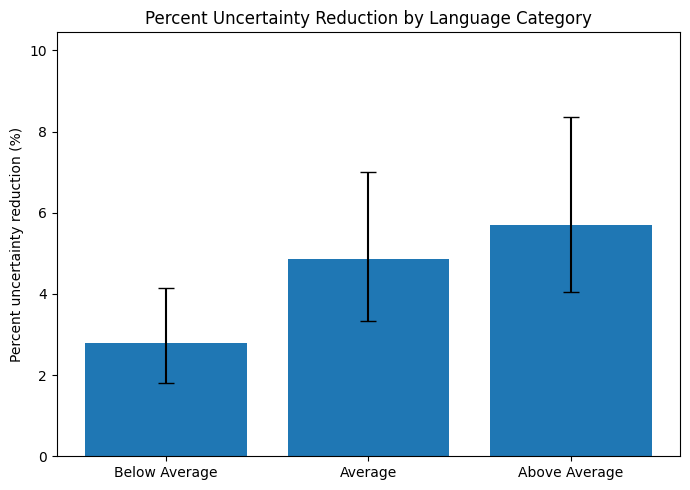}
    }
    \hfill
    \subfloat[Cognitive]{
        \includegraphics[width=0.48\textwidth]{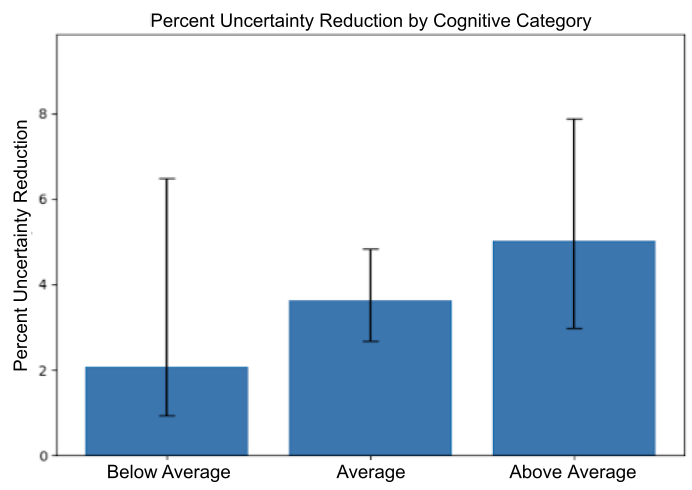}
    }
    \caption{The percent of reduction in uncertainty from knowing the current speaker identity for each Bayley group.}
    \label{fig:mi}
\end{figure*}

\subsection{Hierarchical Bayesian Modeling}

To examine how transition likelihoods changed over the course of a
conversation, we calculated the posterior slope for each transition
$k \to j$ and Bayley group $g$. For each dyad $d$, let $\tau_{d,t}$
denote scaled turn number, with $\tau_{d,t} \in [-1,1]$. The
group-specific slope for transition $k \to j$ is

\begin{equation}
    \beta_{d,k,j,g} =
    \begin{cases}
        \beta_{\mu,k,j} + b_{d,k,j} + \beta_{\mathrm{low},k,j},
        & g = \text{Low Bayley},\\
        \beta_{\mu,k,j} + b_{d,k,j},
        & g = \text{Average Bayley},\\
        \beta_{\mu,k,j} + b_{d,k,j} + \beta_{\mathrm{high},k,j},
        & g = \text{High Bayley},
    \end{cases}
\end{equation}

where $\beta_{\mu,k,j}$ is the population-level slope for the Average
Bayley group and $b_{d,k,j}$ is the dyad-specific random slope. The
time-varying logit for transition $k \to j$ is

\begin{equation}
    \eta_{d,k,j,g}(\tau)
    = \alpha_{d,k,j} + \beta_{d,k,j,g}\tau.
\end{equation}

Because overlap is the reference next-speaker category,
$\eta_{d,k,\mathrm{overlap},g}(\tau)=0$. For
$j \in \{\mathrm{mother},\mathrm{child}\}$, the transition
probabilities are

\begin{equation}
    P_{d,j \mid k,g}(\tau) =
    \frac{\exp(\eta_{d,k,j,g}(\tau))}
    {1 + \sum_{r \in \{\mathrm{mother},\mathrm{child}\}}
    \exp(\eta_{d,k,r,g}(\tau))},
\end{equation}

and

\begin{equation}
    P_{d,\mathrm{overlap} \mid k,g}(\tau) =
    \frac{1}
    {1 + \sum_{r \in \{\mathrm{mother},\mathrm{child}\}}
    \exp(\eta_{d,k,r,g}(\tau))}.
\end{equation}

We calculated the finite-difference slope for each posterior draw as

\begin{equation}
    \frac{P_{d,j \mid k,g}(1) - P_{d,j \mid k,g}(-1)}{2}.
\end{equation}

We then calculated the 95\% Highest Density Interval (HDI) and the mean of each posterior slope, as shown in Figure \ref{fig:mi}. Our findings show that 95\% HDIs for the between-group slope differences include 0. However, the groups differ in whether their HDIs overlap 0 (i.e., whether transition probabilities change over the interaction). 

When analyzing these interactions, we observe distinct patterns of turn-taking across Bayley language score groups. For the low Bayley groups, mothers become less likely to take consecutive turns as the interaction goes on. However, for the low-scoring Bayley group, the shift is toward mother-to-child turn, while the high-scoring Bayley group shifts toward mother to overlapping turns. Additionally, for children in the low and average Bayley groups, when the child is the one speaking, we see that children tend to continue with consecutive turns, whereas overlapping decreases over time across all three groups.
The cognitive Bayley groups follow similar patterns, with slight variations when the mother starts a turn. Again, low Bayley groups are less likely to transition from mother to mother over time; however, in the cognitive Bayley, average Bayley children are also less likely to make this transition, whereas high Bayley children have HDIs overlapping at 0. Both the low and average Bayley groups become more likely to transition from mother to child over time instead. Only the average Bayley children are more likely to transition from child to child over time, although all groups of children become less likely to transition from child to overlapping turns. Finally, high-Bayley children become more likely to transition from overlapping speech to non-overlapping speech over time.   

Developmentally, these findings suggest that mother-child turn-taking changes over the course of an interaction in ways that vary by developmental level, even though differences between the Bayley groups themselves are not statistically credible. Children with lower Bayley scores tended to show increasing child-led responding (i.e., more mother→child and child→child transitions), whereas children with higher language scores showed increasing overlap following maternal turns, potentially reflecting more active participation or attempts to enter the conversation. Across groups, overlapping responses following child speech became less common over time, suggesting increasing conversational organization as interactions progressed. Overall, the results indicate that developmental differences are reflected more in within-group changes in conversational dynamics over time than in clear between-group differences.

\begin{figure*}[t]
    \centering
    \subfloat[Language]{
        \includegraphics[width=0.48\textwidth]{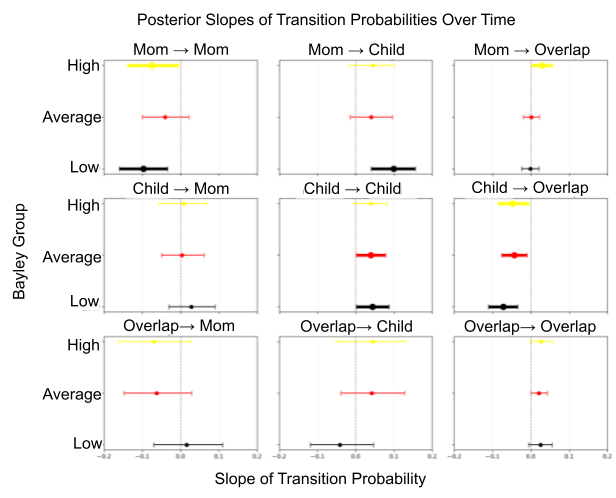}
    }
    \hfill
    \subfloat[Cognitive]{
        \includegraphics[width=0.48\textwidth]{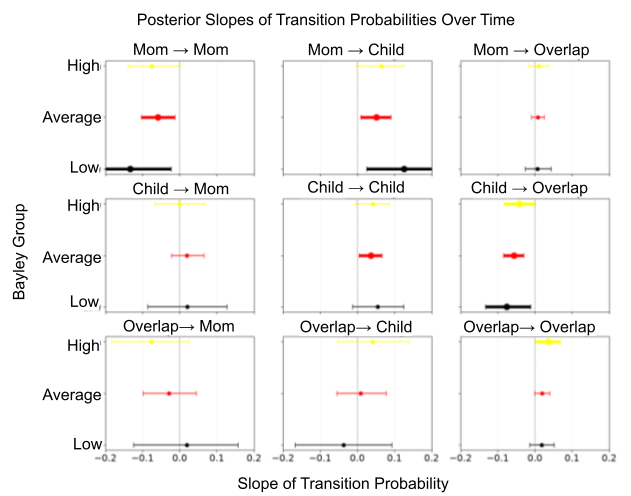}
    }
    \caption{Graph of posterior slopes with HDI ranges for each Bayley group and turn transition type. HDI ranges that do not overlap with a 0 slope (are credibly non-zero trends) are bolded. The bars are colored by Bayley group (yellow$\to$high, red$\to$average, and black$\to$low)}
    \label{fig:mi}
\end{figure*}

\section{Conclusion}
This work investigated how child cognitive and language development is reflected in mother-child conversational dynamics by combining sequence similarity, information theory, and hierarchical Bayesian modeling. Across these complementary approaches, developmental differences were consistently reflected within the organization of turn-taking between the dyad. 

Sequence-similarity analyses found significantly lower similarity in comparisons involving low Bayley scores than in the Average-Average reference comparison; no other group contrasts were significant. Three-turn sequence analyses found Holm-adjusted differences in three sequences across language Bayley groups, but none across cognitive Bayley groups. In language comparisons, AAA adult-dominated turn-taking was most frequent in the low Bayley group, while serve-and-return sequences (ACA and CAC) increased in frequency across groups.

Mutual information tended to show that speaker transitions became increasingly predictable as developmental scores increased. Dyads with above-average language and cognitive scores exhibited approximately twice the uncertainty reduction of below-average dyads, indicating that conversations involving higher-performing children might have followed more structured and dependent turn-taking patterns, however confidence intervals overlapped so no general claims can be made. 

Although the hierarchical Bayesian analysis did not identify statistically credible differences in temporal transition slopes between Bayley groups, it revealed meaningful within-group changes over the course of interactions. Lower-scoring dyads tended to become increasingly child-led over time, with more mother-to-child and child-to-child transitions. In contrast, higher language-scoring dyads showed greater overlap following maternal turns, which could indicate more active conversational participation. Across all groups, overlap following child speech generally decreased over time, suggesting that interactions became more organized as conversations progressed.

Taken together, these findings suggest that early developmental differences are reflected in multiple aspects of mother-child conversation, including interaction structure, predictability, and temporal dynamics. By combining these different methods, we provide a more comprehensive framework for characterizing conversation development. 

Limitations of this work should be considered when interpreting these findings. First, the developmental group differences identified by the mutual information and hierarchical Bayesian models were limited by statistical uncertainty. Second, manual annotation of conversational turns remains a potential source of measurement error. While a second annotator independently labeled a subset of the recordings and achieved high agreement, additional annotation on larger portions of the dataset would further strengthen confidence in turn-taking annotations. Finally, this work included 94 mother-child dyads. Although this is a large sample in the context of child development, future work should extend this workflow to larger longitudinal datasets, to validate findings, examine finer-grained conversational behaviors beyond speaker identity, and investigate the use of these findings in interventions designed to support and encourage high-quality mother-child interactions.


%
\clearpage
\appendices

\begin{figure*}[!t]

\section{}
\label{app:bayesian}

\centering
\includegraphics[
    width=\textwidth,
    height=0.82\textheight,
    keepaspectratio
]{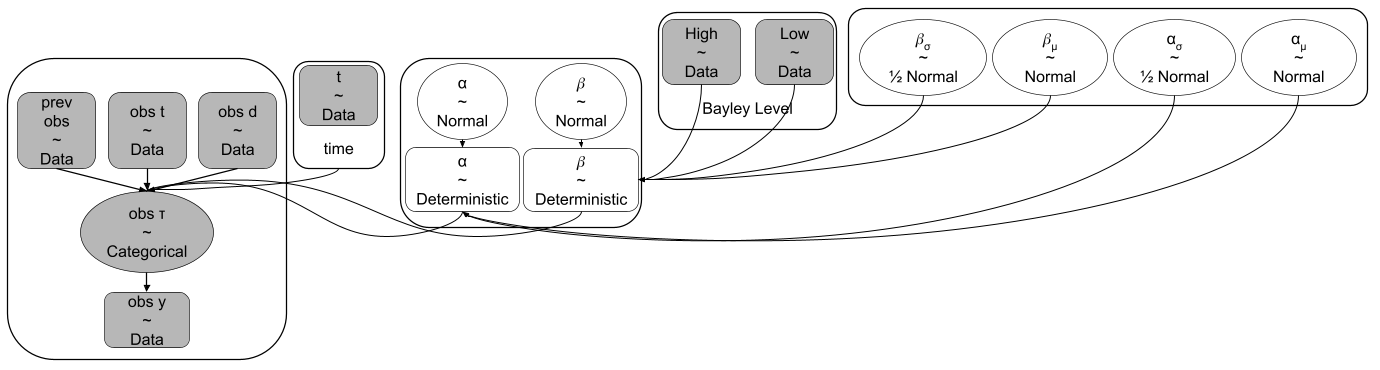}

\caption{Visualization of the Bayesian hierarchical model used.}
\label{fig:bayesian}

\end{figure*}

\begin{table*}[!t]

\section{}

\centering
\begin{minipage}[t]{0.48\textwidth}
\centering
\resizebox{\textwidth}{!}{%
\begin{tabular}{|l|l|l|}
\hline
\textbf{Sequence} & \textbf{ANOVA} &
\textbf{\begin{tabular}[c]{@{}l@{}}Effect Size \\ ($\eta^2$)\end{tabular}} \\ \hline
AAA & F(2.00, 91.00) = 2.26, p = 1.0 & 0.05 \\ \hline
AAC & F(2.00, 91.00) = 2.18, p = 1.0 & 0.05 \\ \hline
AAO & F(2.00, 91.00) = 2.11, p = 1.0 & 0.04 \\ \hline
ACA & F(2.00, 91.00) = 4.52, p = 0.35 & 0.09 \\ \hline
ACC & F(2.00, 91.00) = 1.41, p = 1.0 & 0.03 \\ \hline
ACO & F(2.00, 91.00) = 0.54, p = 1.0 & 0.01 \\ \hline
AOA & F(2.00, 91.00) = 0.35, p = 1.0 & 0.01 \\ \hline
AOC & F(2.00, 91.00) = 0.92, p = 1.0 & 0.02 \\ \hline
AOO & F(2.00, 91.00) = 0.05, p = 1.0 & 0.00 \\ \hline
CAA & F(2.00, 91.00) = 2.91, p = 1.0 & 0.06 \\ \hline
CAC & F(2.00, 91.00) = 4.71, p = 0.31 & 0.09 \\ \hline
CAO & F(2.00, 91.00) = 1.87, p = 1.0 & 0.04 \\ \hline
CCA & F(2.00, 91.00) = 1.25, p = 1.0 & 0.03 \\ \hline
CCC & F(2.00, 91.00) = 0.77, p = 1.0 & 0.02 \\ \hline
CCO & F(2.00, 91.00) = 0.86, p = 1.0 & 0.02 \\ \hline
COA & F(2.00, 91.00) = 0.39, p = 1.0 & 0.01 \\ \hline
COC & F(2.00, 91.00) = 0.17, p = 1.0 & 0.00 \\ \hline
COO & F(2.00, 91.00) = 0.53, p = 1.0 & 0.01 \\ \hline
OAA & F(2.00, 91.00) = 0.49, p = 1.0 & 0.01 \\ \hline
OAC & F(2.00, 91.00) = 0.19, p = 1.0 & 0.00 \\ \hline
OAO & F(2.00, 91.00) = 0.88, p = 1.0 & 0.02 \\ \hline
OCA & F(2.00, 91.00) = 0.05, p = 1.0 & 0.00 \\ \hline
OCC & F(2.00, 91.00) = 0.96, p = 1.0 & 0.02 \\ \hline
OCO & F(2.00, 91.00) = 0.53, p = 1.0 & 0.01 \\ \hline
OOA & F(2.00, 91.00) = 0.06, p = 1.0 & 0.00 \\ \hline
OOC & F(2.00, 91.00) = 0.37, p = 1.0 & 0.01 \\ \hline
OOO & F(2.00, 91.00) = 0.31, p = 1.0 & 0.01 \\ \hline
\end{tabular}%
}

\caption{ANOVA differences in frequency of 3-turn sequences by cognitive Bayley group.}
\label{tab:cognitive_all}

\end{minipage}
\hfill
\begin{minipage}[t]{0.48\textwidth}
\centering
\resizebox{\textwidth}{!}{%
\begin{tabular}{|l|l|l|}
\hline
\textbf{Sequence} & \textbf{ANOVA} &
\textbf{\begin{tabular}[c]{@{}l@{}}Effect Size \\ ($\eta^2$)\end{tabular}} \\ \hline
AAA & F(2.00, 91.00) = 7.47, p = 0.025  & 0.14 \\ \hline
AAC & F(2.00, 91.00) = 4.64, p = 0.28  & 0.09 \\ \hline
AAO & F(2.00, 91.00) = 4.67, p = 0.28  & 0.09 \\ \hline
ACA & F(2.00, 91.00) = 13.06, p < 0.001 & 0.22 \\ \hline
ACC & F(2.00, 91.00) = 3.90, p = 0.48  & 0.08 \\ \hline
ACO & F(2.00, 91.00) = 0.12, p = 1.0  & 0.00 \\ \hline
AOA & F(2.00, 91.00) = 4.44, p = 0.32  & 0.09 \\ \hline
AOC & F(2.00, 91.00) = 0.51, p = 1.0  & 0.01 \\ \hline
AOO & F(2.00, 91.00) = 1.04, p = 1.0  & 0.02 \\ \hline
CAA & F(2.00, 91.00) = 4.13, p = 0.40  & 0.08 \\ \hline
CAC & F(2.00, 91.00) = 8.41, p = 0.01  & 0.16 \\ \hline
CAO & F(2.00, 91.00) = 1.26, p = 1.0  & 0.03 \\ \hline
CCA & F(2.00, 91.00) = 2.56, p = 1.0  & 0.05 \\ \hline
CCC & F(2.00, 91.00) = 0.31, p = 1.0  & 0.01 \\ \hline
CCO & F(2.00, 91.00) = 0.22, p = 1.0  & 0.00 \\ \hline
COA & F(2.00, 91.00) = 0.17, p = 1.0  & 0.00 \\ \hline
COC & F(2.00, 91.00) = 0.63, p = 1.0  & 0.01 \\ \hline
COO & F(2.00, 91.00) = 1.33, p = 1.0  & 0.03 \\ \hline
OAA & F(2.00, 91.00) = 2.13, p = 1.0  & 0.04 \\ \hline
OAC & F(2.00, 91.00) = 1.10, p = 1.0  & 0.02 \\ \hline
OAO & F(2.00, 91.00) = 1.00, p = 1.0  & 0.02 \\ \hline
OCA & F(2.00, 91.00) = 0.45, p = 1.0  & 0.01 \\ \hline
OCC & F(2.00, 91.00) = 0.88, p = 1.0  & 0.02 \\ \hline
OCO & F(2.00, 91.00) = 0.15, p = 1.0  & 0.00 \\ \hline
OOA & F(2.00, 91.00) = 1.44, p = 1.0  & 0.03 \\ \hline
OOC & F(2.00, 91.00) = 0.60, p = 1.0  & 0.01 \\ \hline
OOO & F(2.00, 91.00) = 2.59, p = 1.0  & 0.05 \\ \hline
\end{tabular}%
}

\caption{ANOVA differences in frequency of 3-turn sequences by language Bayley group.}
\label{tab:language_all}

\end{minipage}

\end{table*}

\clearpage
\bibliographystyle{IEEEtran}
\bibliography{bib}
\end{document}